\documentclass[aps,prb,preprint,endfloats]{revtex4}
\makeatletter
\def\@dotsep{4.5}
\makeatother
\usepackage{graphicx}
\usepackage{dcolumn}
\usepackage{bm}

\begin{document}

\title{Models of the World human population growth- critical analysis}

\author{Michael Golosovsky\footnote{electronic mail: golos@cc.huji.ac.il}}
\ \affiliation{The Racah Institute of Physics, The Hebrew University of Jerusalem}

\date{\today} 
\begin{abstract} 
We analyze  mathematical models of the global human population growth and compare them to actual dynamics of the world population and of the world surplus product. We consider a possibility that the so-called world's demographic transition is not a dynamic crossover but a phase transition that affects all aspects of our life.
\end{abstract}

\maketitle

\section{Introduction}
\subsection{Empirical models}
The human population of the Earth $N_{E}$  attracted much attention after publication of the seminal work of  Malthus who  realized that it should exhibit the unlimited exponential growth:
\begin{equation}
\frac{dN_{E}}{dt}=rN_{E}.\label{Malthus}
\end{equation}
The fears were partially dispersed by Verhulst  who introduced the logistic equation, 
\begin{equation}
\frac{dN}{dt}=rN\left(1-\frac{N}{K}\right),\label{logistic}
\end{equation}
to account for the  population dynamics of closed communities. Here, $r$ is the  growth rate and $K$ is the carrying capacity. 
This equation accounts fairly well for the growth of small communities  but it fails to describe the long-time dynamics of the human population of the Earth.  

As it was shown in the seminal work of von Foerster, Mora, and Amiot \cite{Foerster},   the available to them data (up to year 1960) could be fairly well described by the empirical dependence
\begin{equation}
N_{E}(t)=\frac{C}{(t_{c}-t)^{\alpha}}\label{hyperbolic}
\end{equation}
with the parameters: $\alpha\approx 1$,  $C=1.8\cdot 10^{11}$ and $t_{c}=$2026. The corresponding growth rate is:
\begin{equation}
\frac{dN_{E}}{dt}\approx\frac{C}{(t_{c}-t)^{2}}.\label{rate_hyperbolic}
\end{equation}
The most striking feature of Eqs. \ref{hyperbolic},\ref{rate_hyperbolic} is the  divergence of $N_{E}$ and $dN_{E}/dt$ at finite time, $t_{c}$. This indicates that  the above equations are inappropriate in the vicinity of $t_{c}$. Indeed, since 1960, the global human population growth deviated from the hyperbolic dependence indicated by the Eqs.\ref{hyperbolic},\ref{rate_hyperbolic}. In particular,  the growth rate $\frac{1}{N_E}\frac{dN_{E}}{dt}$ achieved its  maximum value of $\sim 2.1\%$ in 1962 and then was steadily decreasing.
This prompted  the search for the functions that approximate the hyperbolic dependence  given by Eqs.\ref{hyperbolic},\ref{rate_hyperbolic} before year 1960 and replace them by  smoother dependences after  1960.  Several empirically found replacements have been suggested,  including hypergeometric \cite{Koronovskii}, overlay of several exponential or logistic curves \cite{Hanson}, hyperexponential \cite{Varfolomeev}, delayed logistic curves \cite{Haberl,Yukalov}, and others. The most insightful empirical approach was suggested by S.P. Kapitza \cite{Kapitza} who modified the Eq.\ref{rate_hyperbolic} as follows:
\begin{equation}
\frac{dN_{E}}{dt}=\frac{C}{(t_{c}-t)^{2}+\tau^{2}}.\label{rate_kapitza}
\end{equation}
Here, $\tau$ is a microscopic time scale for which Kapitza took the lifespan of a generation, $\sim$ 45 years. This modification captures the maximum in the relative growth rate and assumes that the human population eventually comes to saturation. The subsequent studies sought to justify this empirical approach.

\subsection{Mathematical models}
\subsubsection{Models considering the carrying capacity of the Earth}
In order to understand the future trends of the global human population growth, several  non-empirical mathematical models have been developed. These models aimed to derive Eqs.\ref{hyperbolic},\ref{rate_hyperbolic} from the  "first principles". This approach implies that the equations are consequences of some plausible scenario while the  parameters are empirical.  Most of these models \cite{Artzrouni,Cohen,Kremer,Komlos,Podlazov,Galor} quantified the verbal approach of Boserup, Simon, Jones, etc. who attributed the accelerating growth  of the human population of the Earth, $N_{E}$, to  positive feedback between the population size and the Earth's carrying capacity, $K_{E}$.  Then, in addition to  Eq.\ref{logistic} which accounts for the fast growth of the world population, these models introduced an additional equation that accounts for the slow dynamics of the population growth resulting from the gradual increase of the carrying capacity:  
\begin{equation}
\frac{dK_{E}}{dt}=\gamma K_{E}N_{E}\label{K-growth}
\end{equation}
The coefficient $\gamma$ quantifies the rate with which the human race expands the  carrying capacity of the Earth. 

In such a way, these models assume two rates of the population growth. The fast rate,  as derived from  Eq.\ref{logistic}, is $\frac{1}{N}dN/dt\sim r$; while  the slow rate, as derived by the Eq.\ref{K-growth}, is $\frac{1}{K}dK/dt\sim \gamma N$.  As far as $r>>\gamma N$ the instantaneous value of the population size (Eq.\ref{logistic}) is $N\approx K$ (the index $E$ has been omitted). Then Eq.\ref{K-growth} reduces to  
\begin{equation}
\frac{dN}{dt}\approx\gamma N^2.\label{autocatalytic}
\end{equation}
This equation describes an autocatalytic process and its solution
is  given by Eq.\ref{hyperbolic} where $C=\gamma^{-1}$, $t_{c}=t_{i}+1/\gamma N_{i}$, and $t_{i},N{i}$ are  initial conditions. Upon approaching $t_{c}$, the population size $N$ increases and the distinction between the slow and fast dynamics eventually disappears. In the limit $\gamma N>>r$, the Eqs.\ref{logistic},\ref{K-growth} yield exponential population growth, $N\propto e^{rt}$. 

At present we do not know whether the human population  will come to saturation in future or will grow continuously, although we want to believe that its growth will be somehow limited.  The Kapitza's conjecture  consists in replacing the Eq.\ref{autocatalytic} by the empirical equation
\begin{equation}
\frac{dN}{dt}=\frac{r^2}{\gamma}\sin^{2}\frac{\gamma N}{r}\label{kapitza1}
\end{equation}
that describes accelerating and then decelerating population growth, whereas the population size is eventually stabilized at  $N_{\infty}=\frac{\pi r}{\gamma}$. The Eq.\ref{kapitza1} describes the dynamic crossover. It operates with the minimal number of parameters -$r$ and $\gamma$, and is mathematically appealing. However, it can't be easily justified and the reasons for the maximum of the growth rate and for the stabilization of the population size remain obscure. 

\subsubsection{Models based on the Gross Domestic Product-GDP}

The carrying capacity, being the important parameter of the demographic models, can not be measured directly. The model  that does not consider explicitly the carrying capacity was developed by  M. Kremer \cite{Kremer} who considered the gross domestic product, $GDP =N(S+m)$, as the key parameter that determines the slow dynamics of the population growth. Here, $N$ is the population size, $m$ is the subsistence level, and $S$ is the surplus product. Kremer related the GDP to the level of technological development $T$,  as follows: $GDP\propto N^{\phi_{1}}T^{\phi_{2}}$, where $\phi_1,\phi_2$ are the exponents that should be found empirically. In fact, Kremer put onto quantitative language the  verbal approach that had been developed earlier by Kuznets, Boserup, Jones, etc. The key assumption of the Kremer's model is that the growth of GDP is spurred by the technology growth. 

The original model developed by Kremer is static while Korotaev, Malkov, and Khaltourina  \cite{Korotaev}  developed a family of  dynamic models basing on Kremer's ideas. In the framework of these models, the  dynamic variable that quantifies the technological development is the surplus product, $S$.  The simplest model considered by Ref.\cite{Korotaev} consists of two equations
\begin{equation}
\frac{dN}{dt}=rNS/m\label{Korotaev1a}
\end{equation}
\begin{equation}
\frac{dS}{dt}=\gamma NS\label{Korotaev1b}
\end{equation}
with two empirical parameters: $r$ is the rate of the population growth, and $\gamma$ has now the meaning of the average creative ability of a person. The parameter $m$ characterizes the scale of the surplus product $S$ (it can be chosen to be equal to subsistence level) and it has been introduced here to be consistent with the notation of the Eq.\ref{Malthus}. 

In such a way, the Eq. \ref{Korotaev1a} is the modification of the Eq.\ref{Malthus}, while  Eq.\ref{Korotaev1b} captures the  Kremer's idea. The relation to Kremer's work is even more evident if we notice that for  $\phi_{1}\sim 1$, the definition of the technological level  by Kremer :$T\propto (GDP/N)^{1/\phi_{2}}$, is closely related to the definition of the surplus product: $(S+m)\propto GDP/N$. 

The relation between $N$ and $S$ can be found by  dividing Eq.\ref{Korotaev1a} by Eq.\ref{Korotaev1b}. This yields $N\propto S$. In such a way, the Eqs.\ref{Korotaev1a},\ref{Korotaev1b} describe the positive feedback between the surplus product and the population growth - from the one hand; and the positive feedback between the increasing population and the growth of the surplus product, from another hand \cite{KKM-problem}. The solution of these equations exhibits finite-time singularity for both $N(t)$ and $S(t)$.

 The field of applicability of Eqs.\ref{Korotaev1a},\ref{Korotaev1b} is evident from their very structure: the right side looks as if it were the first term of the power series in the small parameter $S/m$. In other words,   Eqs.\ref{Korotaev1a},\ref{Korotaev1b} assume that $S/m<<1$, i.e. they should describe the period before the year 1870 when $S/m\sim 1$. One can go beyond this approximation and modify the Eqs.\ref{Korotaev1a},\ref{Korotaev1b} to extend their applicability range to beyond the year 1870.  Indeed, if we assume that the surplus product goes to creation of new working places, the relation of carrying capacity to surplus product is especially simple: $K=GDP/m=N(S/m+1)$. We replace Eq.\ref{Korotaev1a}  by the logistic equation where $K$ is expressed through $S$ and find
\begin{equation}
\frac{dN}{dt}=\frac{rNS}{S+m}\label{Korotaev2a}
\end{equation}
This equation  introduces negative feedback between the population growth and the growing population. This has some stabilizing effect and consequently, the solution of Eqs. \ref{Korotaev2a},\ref{Korotaev1b} does not diverge. In the long run, the growth rate of $N$  comes to saturation, while the growth rate of $S$ is unlimited. The relation between $N$ and $S$ can be found by dividing Eq.\ref{Korotaev1b} by Eq.\ref{Korotaev2a}. This yields $\frac{dN}{dt}=\frac{r}{\gamma(S+m)} \frac{dS}{dt}$. The solution of this equation is 
\begin{equation}
S(t)=m\left(e^{\frac{N(t)}{N_{0}}}-1\right)\label{S_N}
\end{equation}
where $N_{0}=r/\gamma$.

In what follows we analyze the data on the relative growth rates: $r_{N}=\frac{dN}{dt}:N$ and $r_{S}=\frac{dS}{dt}:S$ and compare them to the prediction of Eqs.\ref{Korotaev1a},\ref {Korotaev2a},\ref{S_N}. Figure \ref{fig:S_exp} shows that the relation between $N$ and $S$ is superlinear and is fairly well described  by Eq.\ref{S_N}, with $N_{0}=1.65\cdot 10^9$, the agreement breaks only for $N>3\cdot 10^9$ (this corresponds to years 1960-1970). Figure  \ref{fig:r_N} shows that the human population growth rate linearly increases with $N$, achieves its maximum value of $r_{N}^{max}=0.021$ at $N\approx 3\cdot 10^9$ and then starts to decrease.  The Eq.\ref{Korotaev1b} correctly predicts dynamics of $r_{N}$ at $N<2\cdot 10^9$ and deviates from the actual data at higher $N$. Figure  \ref{fig:r_S} shows that the World surplus product growth rate linearly increases with $N$, achieves its maximum value of $r_{S}^{max}=0.04$ at $N\approx 3\cdot 10^9$ and then starts to decrease. The Eq.\ref{Korotaev1b} correctly predicts dynamics of $r_{S}$ at $N<2\cdot 10^9$ an deviates from the actual data at higher $N$. We conclude that the Eqs.\ref{Korotaev1b},\ref{Korotaev2a} extend the range of applicability of the Korotaev, Malkov, and Khaltourina model \cite{Korotaev} from the year $\sim$ 1870 to the year  $\sim$ 1960. Since no new parameters/variables have been introduced, this extension  belongs to the same  family of models \cite{Korotaev}. 

Figure \ref{fig:r_S_S} shows population growth rate $r_{N}$ and the surplus product growth rate $r_{S}$ versus surplus product, $S/m$. Both  $r_{N}$ and $r_{S}$ grow with $S$, achieve the maximum  at  $S/m\sim 5-6$ (this corresponds to $N\sim 3\cdot 10^9$) and then decrease. The model correctly predicts initial increase of  $r_{N}$ and $r_{S}$ with $S$ (that corresponds to $S/m\leq 1$) while pronounced deviations from the model occur for $S/m\geq 2$.


\subsubsection{Models accounting for the demographic transition}
The maximum in the population growth rate (Fig.\ref{fig:r_N}) is usually associated with the  "demographic transition" \cite{Chesnais} which results from  the decrease of mortality rate  and the subsequent birth rate decrease. Detailed statistical studies indicate that the reason for decreasing population growth in developed countries nowadays is the declining birth rate \cite{Chesnais} whereas there is a strong  anticorrelation between the level of woman education and fertility. Korotaev, Khaltourina and Malkov \cite{Korotaev} captured this  by introducing an  additional dynamic variable: the fraction of literate population, $l$. They modified the Malthus equation,  Eq.\ref{Korotaev1a}, to account for the negative feedback between the population growth and the literacy level:
\begin{equation} 
\frac{dN}{dt}=rNS(1-l).\label{KKM_N}
\end{equation}
The dynamics of the surplus product remained the same (Eq.\ref{Korotaev1b}) while the dynamics  of $l$ has been described by the following equation,
\begin{equation}
\frac{dl}{dt}=aSl(1-l),\label{literacy}
\end{equation}
where $a$ is a new empirical parameter. 


When the initial educational level of the population is low, this model predicts accelerating growth of $N$, $S$, and $l$. Eventually, when $l$ comes to saturation, $N$ also achieves saturation while  $S$ does not saturate and grows exponentially. Therefore, this extended  model captures the nonmonotonous dependence of $r_{N}$ on $N$ (Fig.\ref{fig:r_N}) but fails to account for the nonmonotonous dependence of $r_{S}$ on $N$ (Fig.\ref{fig:r_S}).  

\subsubsection{Critical assessment of the above models}
The common feature of all previously discussed models is that they describe the growth of the World human population growth, GDP, surplus product, literacy, etc. by using  ordinary differential equations containing first-order time derivatives. In the framework of these models, the nonmonotonous dependence of the growth rate on time (demographic transition)  results from the dynamic crossover. This means that at all times there are several  factors  affecting  population growth and these factors operate \emph{simultaneously}. At small population number, $N<3\cdot 10^9$, one factor wins and the growth rate increases with $N$; while for high population number, $N>3\cdot 10^9$, another factor wins and the growth rate decreases with $N$. When $N\sim 3\cdot 10^9$ the gradual transition from one regime to another occurs. 

Several features in actual data challenge this picture. First, the transition from increasing to decreasing  trend in $r_{N}$ vs $N$ dependence is  very sharp (Fig.\ref{fig:r_N}). Second, it is not clear why transitions in $r_{N}$  and $r_{S}$ occur simultaneously, in 1960-1970 and at the same value of $N\sim 3\cdot 10^9$ (Fig.\ref{fig:r_S_S}). Other parameters also undergo especially fast change in the same time period- 1960-1970;  these include age structure of the population, the level of literacy, urbanization \cite{Korotaev}, financial indices \cite{Sornette} etc. All these features are more naturally accounted for from another perspective-  phase transition. While the commonly accepted approaches \cite{Korotaev} focus on time-dependent, dynamic properties of the population growth, the phase transition approach focusses on how demographic and economic variables depend on control parameters such as population or surplus product. 

\section{Demographic transition as a Phase transition}
The notion of  phase transition has been developed in the context of condensed matter physics. In the system of many interacting particles/agents, when the control parameter (temperature, pressure, density, etc.) varies, the  system can progress  abruptly from the disordered phase  where the radius of correlations is finite, to the ordered phase, which is characterized by the long-range order. This situation can be usually described using the order parameter which is zero in the disordered phase and non-zero in the ordered phase.  The properties of the system as a function of the control parameter are frequently non-analytic at the transition point, in particular, the correlation length diverges upon approaching the phase transition and becomes infinite in the ordered phase. Divergence of many physical properties at the phase transition is related to the divergence of the correlation length.  The dynamic properties also undergo dramatic changes and the fluctuations  grow on the both sides of the phase transition \cite{Stanley}. The difference between the phase transition and crossover scenario is in the following: for the former, the ordered state is characterized by an emerging new property- the order parameter which was absent in the disordered state; while for the latter scenario all factors were present in both states.

Phase transitions in social systems (such as financial markets, traffic flow, etc.) were noticed long ago  \cite{Montroll,Stauffer}. The extrapolated divergence of the human population growth rate at years 2026-2040  has been also interpreted as some kind of transition \cite{Foerster,Sornette,Korotaev}. However, the growth rate divergence appears only in the "mean-field" models. More realistic models lift the divergence and yield earlier date for the  population growth switch from one regime to another. This probably implies that the phase transition has already taken place in 1960-1970, rather than to occur in 2026-2040. Then the demographic transition of 1960-1970  is not a purely demographic phenomenon but it is  one of the signatures of the global phase transition that has been affecting all aspects of human life. 

It is instructive to discuss the properties of this  transition of 1960-1970 in the context of such a  generic phase transition as lattice percolation \cite{percolation}. Here, the disordered state contains disconnected finite-size clusters  while the ordered state is characterized by the appearance of the infinite cluster that ensures connectivity of the whole system. This analogy prompts us to consider globalization as a hallmark of the phase transition of 1960-1970. One of the most prominent aspects of  globalization is the economic integration. We consider a very crude indicator of the economic integration - the growth of the European Union, in particular we focus on $\eta=N_{EU}/N_{Europe}$ -the fraction of the European population in the states  belonging to European Union (or to its predecessors  such as Common Market that appeared in 1957).   Figure \ref{fig:EC} shows dynamics of $\eta$. Amusingly, this very simple measure of the European integration mimics the World Globalization index as determined by A. Dreher \cite{Dreher} using weighted economic, political and cultural indicators.

Figure \ref{fig:EC}  shows that the onset of European integration can be attributed to the same period- 1960-1970 when the global demographic transition occurred. Serrano \cite{Serrano} came earlier to the same conclusion by considering the historical dynamics of bilateral trade balance. Dependence $\eta(t)$ is very similar to behavior of order parameter at the percolation phase transition \cite{Stauffer}.

If we adopt the hypothesis of  the humanity as a system of interacting agents that undergoes instability/phase transition peaked at 1960-1970 - this raises many interesting questions that prompt extensive scientific research. 
\begin{enumerate}
    \item What is the nature of the phase transition? What is the difference between two phases? Probably, in the years preceding 1960-1970, the most part of the surplus product eventually went to the population growth. However, after 1960-1970 the surplus product was spent more on the increase of the quality of living, i.e. it was channelled to the increase of subsistence level. 
    \item  What is the proper control parameter  that drives this phase transition? Is it the total human population, or average population density, or surplus product, or something else?
    \item   What is the order parameter? Globalization index?
Another candidate for the order parameter could be some measure of information since the appearance of the global information network (Refs.\cite{Kapitza,Naidenov}) after 1970 was very prominent.
    \item  Which parameters diverge upon approaching the transition? What are the critical indices? Besides urbanization and literacy that were studied in Ref.\cite{Korotaev}, it would be interesting to consider the historical dynamics of the warfare indicators such as the weapon range, the power of the explosives, etc. 
    \item  How one can define the correlation length? The possible candidates could be the average city population, or the average size of a polity \cite{Taagepera}.
    \item  What is the statistics of fluctuations at this transition? It is well-known that the fluctuations grow upon approaching the phase transition from both sides. The growth of fluctuations in the context of global population growth has been already noticed \cite{Sornette}. In this context, it would be especially interesting to consider the timeline of the financial crises.
 \item  What is the behavior of the \emph{dynamic} properties of the global human population at the transition? The analog of conductivity of the percolating network would be the signal propagation rate or the rate of adoption of technological innovations. How did these parameters change through historical time?
\end{enumerate}    

\section{Spatially-inhomogeneous and discrete models}
Our capability to introduce innovations that enlarge carrying capacity of the Earth (autocatalycity) translates into human population dynamics equations as a positive feedback. It is well known that population dynamics that includes positive feedback and diffusion leads  to strongly spatially-inhomogeneous population pattern \cite{Shnerb} (for example, desert oases, vegetation patches in arid zones, etc.) and favors agglomeration. Indeed, the increasing economic returns and increasing innovation rate arising from the population agglomeration in cities is well documented \cite{Bettencourt}. This means that in the context of the human population dynamics the spatial-inhomogeneity by itself has autocatalytic properties. Therefore, the models of the human population dynamics should properly address the spatial dimension. 

Note also that the  previously discussed  models were based on ordinary differential equations and didn't take into account neither spatial distribution of the population, nor the \emph{discrete} nature of humans. Very often when  the continuum equations describing population dynamics assume spatially homogeneous population and  predict a very slow growth or even population extinction; the individuals self-organize in spatio-temporally localized adaptive patches which insure their survival and development. In other words, continuum differential equations may fail in predicting the population dynamics of the discrete proliferating  agents \cite{AB}. An interesting example of such approach is the recent study of Yaari et al.\cite{Yaari} of the economics development in Poland after 1990. The Ref.\cite{Yaari} showed that the economics growth was led by few singular "growth centers" these were associated with the University centers. Probably, this shows in a different way the ultimate relation between the education level/literacy and the human population dynamics-see Ref.\cite{Korotaev}.

All this calls for the new generation of the models describing the World human population growth. These models should be  discrete and spatially-dependent.

\section{physical meaning of the parameters meaning of the dynamic models}
We consider here a different topic that arises in relation to dynamical models of the human population growth.  The Eqs.\ref{Korotaev1b},\ref{Korotaev2a} contain two empirical parameters: $r,\gamma$ that should be  somehow related to the human nature. The meaning of the parameter $r$ is more or less clear- it is the relaxation rate of the population to sudden changes. It is determined by the difference in birth rate and mortality and, to the best of our knowledge,  does not exceed  $r^{record}=0.14$. Comparison of the growth rate to the models (Fig.\ref{fig:r_N}) yields $r=0.013\sim 0.1 r_{record}$ that is quite reasonable. Note, that $r$ can be measured from the transient phenomena, for example, how fast the population size recovers from dramatic disaster such as WWII. This yields  $r\sim 0.02-0.03$ that is comparable to $r=0.013$ determined from the Fig.\ref{fig:r_N}.

The meaning of the parameter $\gamma$ is more elusive. Kapitza suggests that $\gamma=1/rU^{2}$ where $U\sim$67,000 is the coherent population unit. This  implies a paradoxical conclusion that \emph{any} coherent population unit consisting of $\sim$ 67,000 individuals will develop into civilization consisting of billions of individuals. To get more clear insight into this paradox we compare the humans and the beavers. Besides their short lifespan (16-20 years), the beavers remind humans in several aspects: they are monogamous, they live in colonies, and most important- by building the dams they shape the landscape according to their needs, i.e. they can expand the carrying capacity. However, the difference between the "civilization" of beavers and human civilization is too obvious. Hence, we seek for  deeper relation between the parameter $\gamma$ and the human nature. 

Korotaev et al.\cite{Korotaev} related the $\gamma$ to the average creativity of a person.  To elaborate further on this subject we assume the following scenario. Human population $N$ adjusts to the current carrying capacity very fast. Carrying capacity $K$ slowly increases due to technological innovations. So far, the spreading of the technological innovations has been the bottleneck that determined the dynamics of the carrying capacity growth.
Equation \ref{K-growth} can be recast as follows:
\begin{equation}
\frac{dK}{K}=\gamma Ndt=\gamma \tau dN_{t}\label{dK}
\end{equation}
where $N_{t}$ is the total number of humans lived on Earth till time $t$, and $\tau$ is the average lifetime. The solution of Eq.\ref{dK} is $K=K_{i}e^{\gamma\tau N_{t}}$  where $K_{i}$ is the carrying capacity at time $t_{i}$ and $N_{t}$ is the total number  of humans that lived on Earth between $t_{i}$ and $t$. Then
\begin{equation}
\gamma=\frac{1}{\tau K}\frac{dK}{dN_{t}} 
=\tau^{-1}\frac{d\ln K}{dN_{t}}\label{gamma}
\end{equation}
For the hyperbolic population growth (Eq.\ref{hyperbolic}), the total number of people lived between $t_{i}$ and $t$ depends logarithmically on time: 
\begin{equation}
N_{total}=-C\ln\frac{t_{c}-t}{t_{c}-t_{i}}\label{N_total}
\end{equation}


The technological innovations are created by people and they are accumulating. This means that the carrying capacity at time $t$ is the result of activity of all  people  that lived before \cite{Cohen}. Therefore, the parameter  $\gamma$ characterizes the average contribution of an individuum to the expansion of the carrying capacity of the Earth.  This  may be interpreted in two ways.
\begin{enumerate}
\item[1] Each individuum contributes to the growth of the Earth carrying capacity in such a way that the \emph{average personal contribution} is $\Delta K=\gamma\tau K$.
\item[2] The carrying capacity increases by giant steps, $\Delta K\sim K$, due to stepwise development of science \cite{V_Turchin} and to scientific/technological revolutions \cite{Kuhn}. Then $\gamma\tau$ is the probability/frequency of these \emph{technological revolutions}. These revolutions  are rare events that trigger a series of smaller innovations which become embodied long before the next revolution occurs.  According to this interpretation, the parameter $\gamma\tau$ is the probability that an inventor or group of inventors makes a major technological/social/administrative breakthrough. According to this scenario, the human population growth is a series of logistic curves, each corresponding to a technological revolution. The quantity  $N_{0}\sim 1/\gamma\tau$ (see Eq.\ref{S_N}) has the meaning of the number of people lived that ensure one major technological revolution.
\end{enumerate}


We believe that the second scenario is more adequate. It has several implications:

\begin{itemize}
\item[*] Log-periodic oscillations around hyperbolic law given by Eq.\ref{hyperbolic} which were noticed by several groups \cite{Sornette,Korotaev} and attributed to cycles, corresponds to a major technological revolution. 
\item[*]The current demographic transition and deviation from the hyperbolic law appear when the technological revolutions occur so frequently that the full potential of the preceding revolution has not been fully realized before the next one occurs. 

\item[*] While the motivation for technological innovations so far was the drive towards increasing carrying capacity, now something changed and the stream of innovations results in increased quality of living  rather than in increasing number of living persons. (This is probably equivalent to increasing subsistence level $m$). Hence, the population growth is not so fast.

\item[*] Is it possible that the very small probability $\gamma\tau\sim 10^{-9}$ is somehow related to the frequency of genetic mutations which is also exceedingly small ($10^{-7}-10^{-8}$)? 


\item [*]  The observation that  bigger populations develop fast, while isolated continents, archipelagos and islands develop slower may be explained quite naturally. This should be related to the probability of appearance of rare events and innovators, and to the discreteness of the population.
\end{itemize}

The link between the above description and that of Ref. \cite {Kapitza} is provided by the discrete character of humans. Indeed, to initiate the positive feedback loop given by Eq.\ref{K-growth}, for the initial group of hominids to expand, it  should  create at least one working place in the lifetime of one generation. This brings us to the minimal group size of $N_{i}=(\gamma\tau)^{-1/2}\sim$ 67,000. 

Another consequence of the approach based on the number of people lived in the certain time interval, is the meaning of "historical time". It has been already noticed \cite{Kapitza} that  with respect to the frequency of historical events, the natural time scale is logarithmic rather than linear. Since the total number of humans that lived on Earth also depends logarithmically on time (Eq.\ref{N_total}), then $N_{t}$ seems to be the "internal clock" of humanity. This conjecture provides the basis for quantitative comparison of the historical development of different isolated communities. According to this interpretation, the internal clock of a community is  the total number of people that ever lived in this community.
\begin{acknowledgments}
I am grateful to Sorin Solomon and Andrey Korotaev  for encouragement  
and for stimulating discussions.
\end{acknowledgments}

 \pagebreak
 
\section{FIGURE CAPTIONS}  
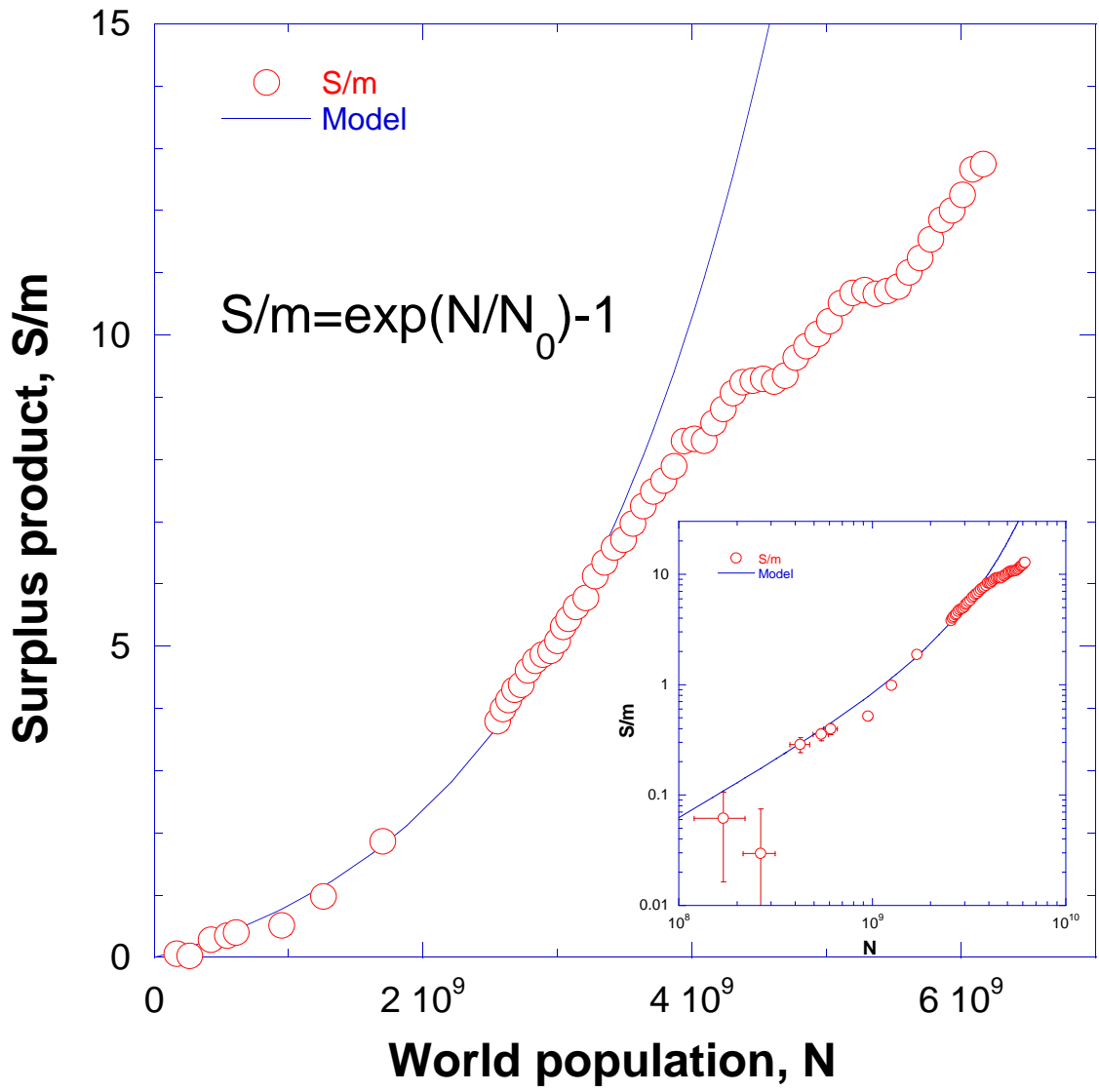
\begin{figure}[ht]
\caption{{Relation between the surplus product $S/m$  and the World human population in the same year. The red circles show the data taken from the US Census database and Refs. \cite{Kremer,Maddison}. The subsistence level is $m=440$ USD. The blue line is the prediction of  Eq.\ref{S_N} with $N_{0}=1.65\cdot 10^9$. }}\label{fig:S_exp}
\end{figure}
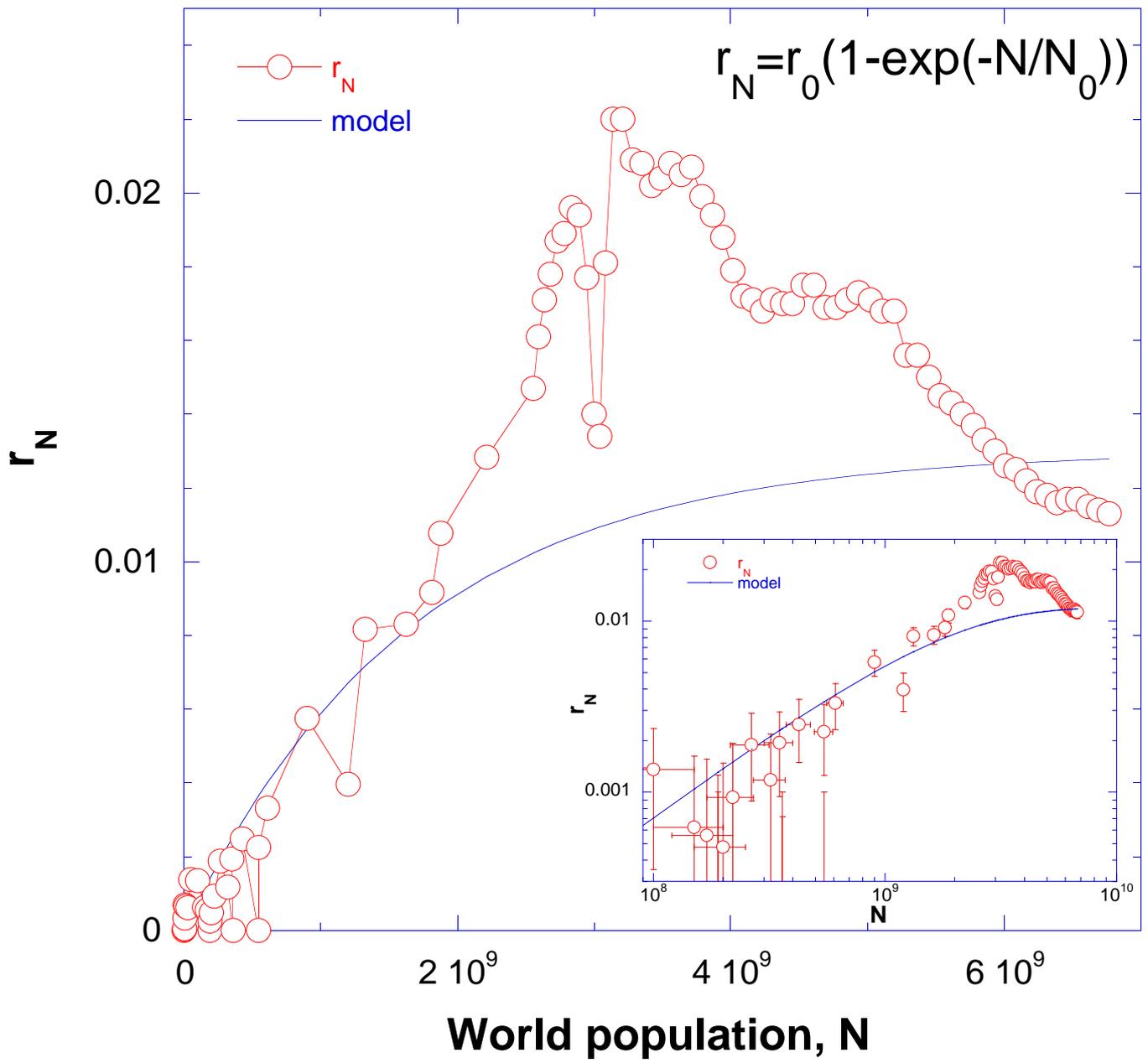
 \begin{figure}[h!]
\caption{{World population growth rate, $r_{N}=\frac{dN}{dt}:N$. The red circles show the data taken from the US Census database and Refs.\cite{Kremer}. The blue line is the prediction of the Eq.\ref{S_N} with $N_{0}=1.65\cdot 10^9$ and $r=0.013$.}}\label{fig:r_N}
\end{figure}
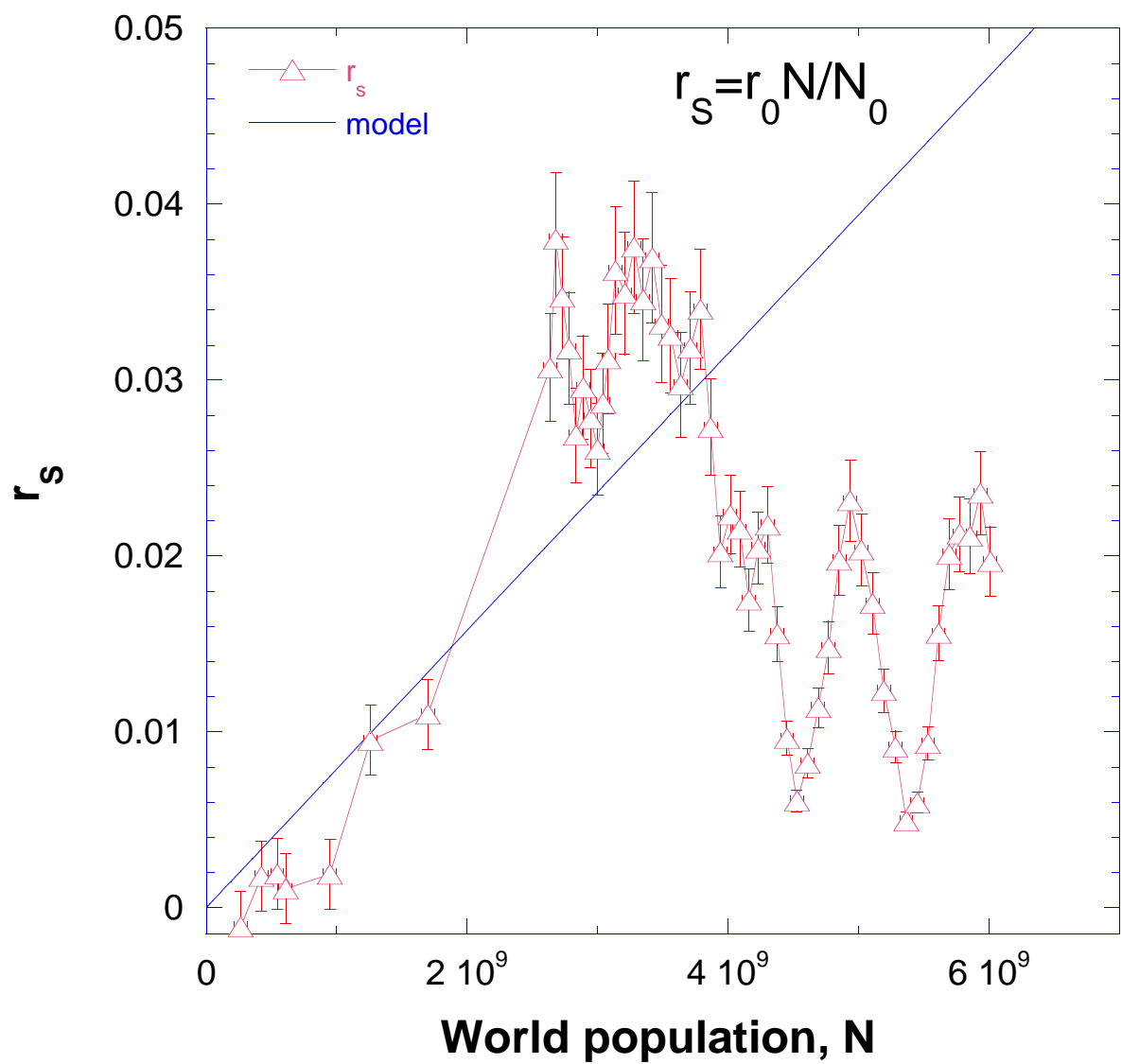
\begin{figure}[h!]
\caption{{World surplus product growth rate, $r_{S}=\frac{dS}{dt}:S$. The red circles show the data taken from the Ref.\cite{Maddison}. The subsistence level is $m=440$ USD. The data were averaged over the five-year interval. The blue line is the prediction of the Eq.\ref{Korotaev2a} with $N_{0}=1.65\cdot 10^9$.}}\label{fig:r_S}
\end{figure}
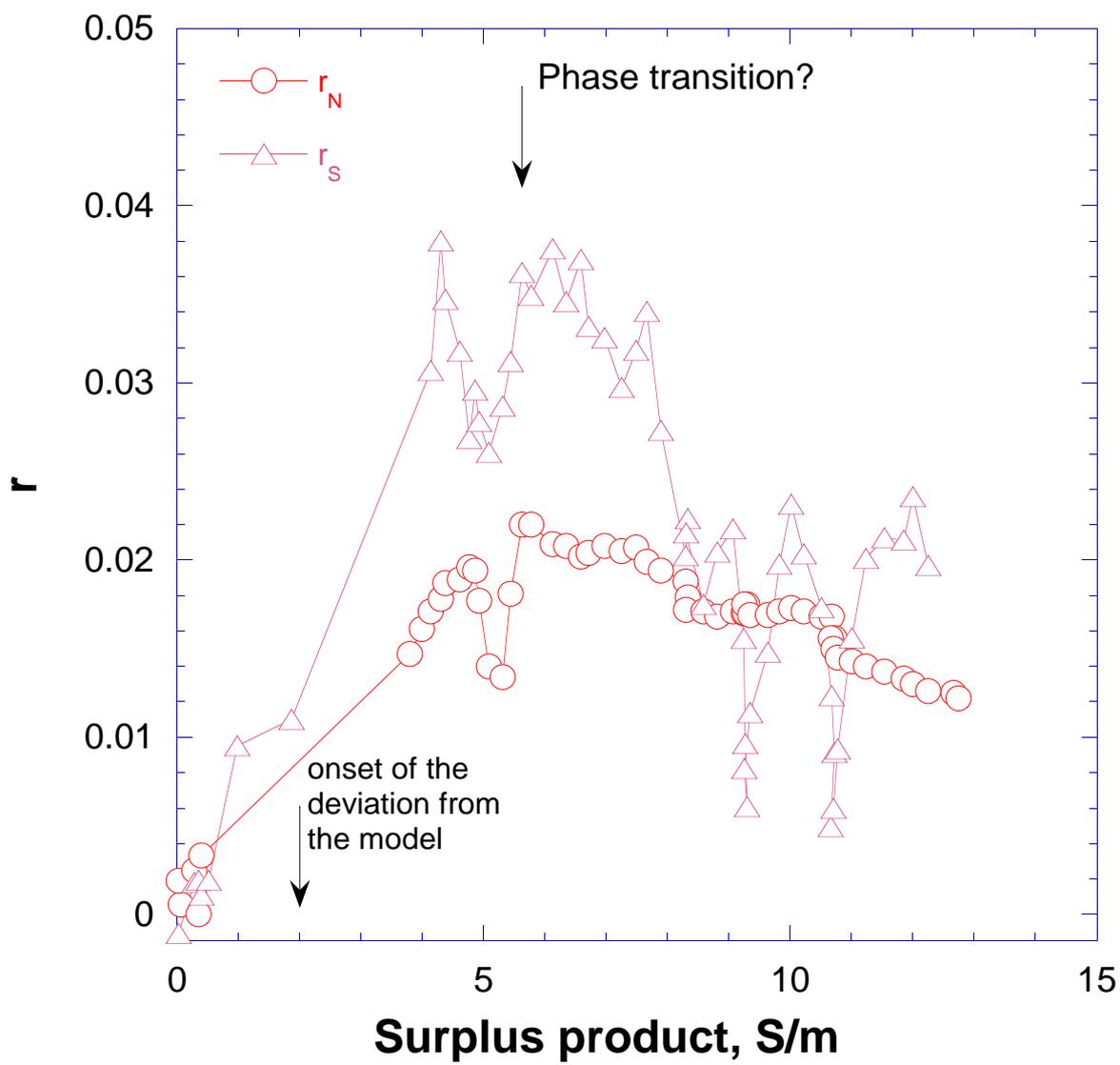
\begin{figure}[h!]
\caption{{Comparison of the World population  and the Surplus product growth rates. The data for $r_{S}$ were averaged over the five-year interval. Note that maximum  $r_{N}$ and maximum $r_{S}$ are achieved simultaneously, at the same value of $S/m$.}}\label{fig:r_S_S}
\end{figure}
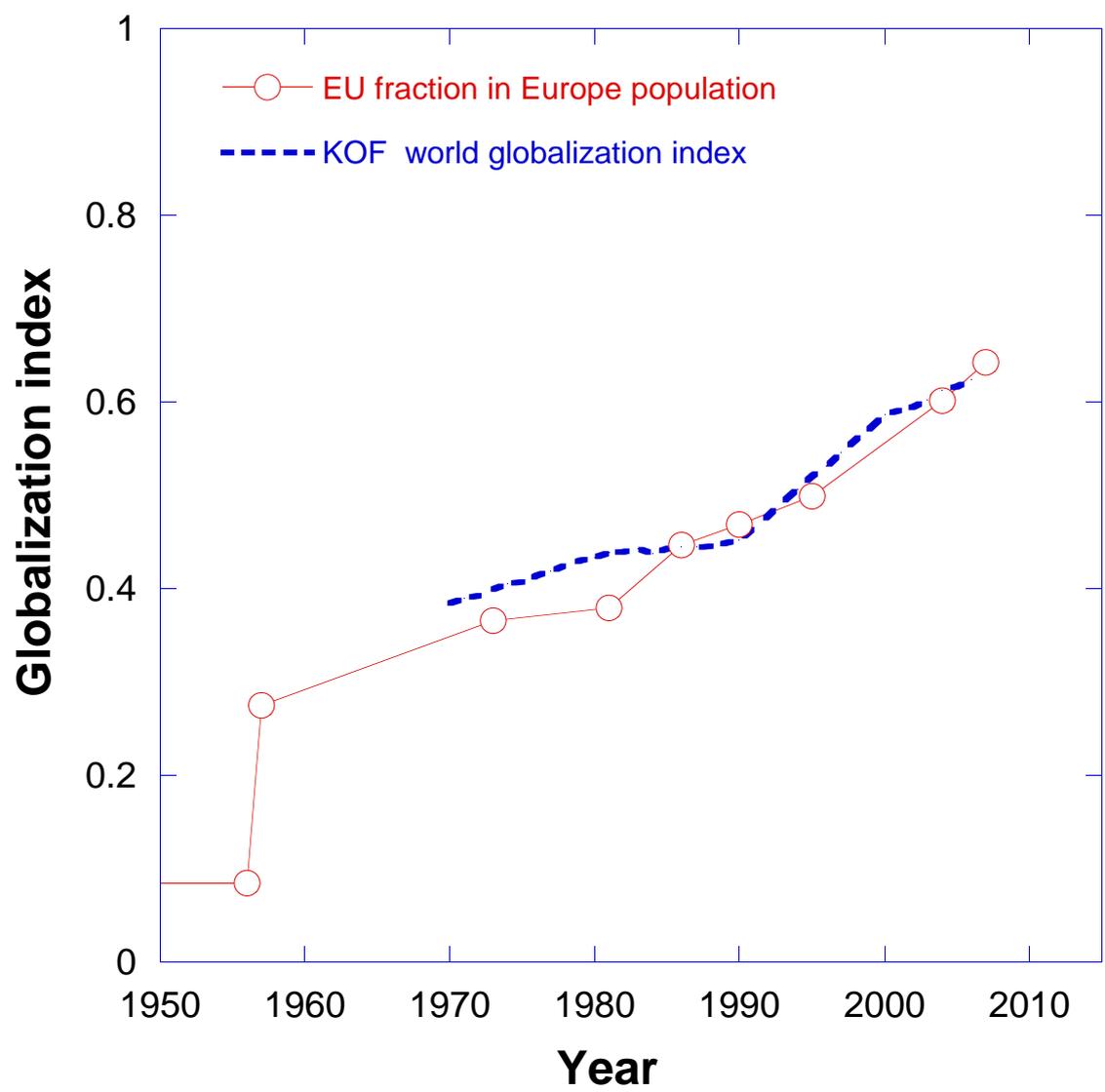
\begin{figure}[h!]
\caption{{Dynamics of the European Union growth, $\eta=N_{EU}/N_{Europe}$ (red circles).  Here, $N_{EU}$ is the population in the states that belong to European Union or to its predecessors (Common Market), while  $N_{Europe}$ is the total European population. The blue dashed line shows KOF globalization index \cite{Dreher}. Note abrupt growth of $\eta$ around 1960 which is followed by slower steady growth afterwards. This is a characteristic behavior of the order parameter at phase transitions. }}\label{fig:EC}
\end{figure}
 \end{document}